# Automated Instrumentation for the Determination of the High-Temperature Thermoelectric Figure-of-Merit

Ashutosh Patel[*] and Sudhir K. Pandey

*School of Engineering, Indian Institute of Technology Mandi,*

*Kamand 175005, Himachal Pradesh, India*

Corresponding author Electronic mail: ashutoshpatelm@gmail.com

## Abstract

In this work, we report the fabrication of a high temperature measurement setup to measure *Figure of merit* ($Z\bar{T}$). This setup facilitates the simultaneous measurement of Seebeck coefficient ($\alpha$), thermal conductivity ($\kappa$), and electrical resistivity ($\rho$) required to calculate $Z\bar{T}$. Measurement of temperature, as well as voltages using same thermocouples, simplified the design of the setup by minimizing sensors and wires. Limited components used in the sample holder further simplify the design and make it small in size and lightweight. The dedicated thin heater is made, which minimizes the heat loss. Further, low heat loss is achieved by optimizing the insulator dimension. To measure power delivered to the heater, 4-wire technique is used. Low cost and commonly available materials used in the fabrication of various components make it more accessible to the user as any parts can be easily replaced in case of any damage occurs. A dedicated program is built in the Python programming language to automate the whole measurement process. *p-type $Bi_{0.36}Sb_{1.45}Te_3$* sample is used to calibrate this measurement setup. The data collected are found to be in good agreement with the reported data.

# INTRODUCTION

The increasing demand of the electricity and decreasing the fossil fuel reserve, forcing us to find renewable energy sources. Due to the poor conversion efficiency of heat engine, a large amount of energy is directly released to the environment as heat. This waste heat can be converted into electricity directly using thermoelectric generators. The main advantages of these generators are small in size, no moving part, low maintenance and suitable for small power generation.[1–3] The major challenges in this area are to develop low cost thermoelectric materials. Currently, bismuth telluride based thermoelectric materials are used, which have good $Z\bar{T}$ at low temperature (~400K).[2] The major potential sites for power generation are available at intermediate temperature (~800K) and high temperature (~1200K) region. Searching of efficient thermoelectric materials for these temperature regions requires a dedicated instrument to characterize samples.

There are two methods commonly used to measure $Z\bar{T}$ of thermoelectric materials. One method gives $Z\bar{T}$ directly and known as Harman method.[4] This method is used to measure $\rho(T)$ and $Z\bar{T}$ over small temperature difference.[4,5] The drawbacks of this method are that it is valid only for small temperature difference and requires adiabatic boundary condition, which is difficult to satisfy at high temperature.[6] In another method, instead of direct measurement of *figure of merit* ($Z\bar{T}$), we measure three parameters, which are Seebeck coefficient ($\alpha$), thermal conductivity ($\kappa$) and electrical resistivity ($\rho$). Figure of merit ($Z\bar{T}$) is calculated using equation,

$$Z\bar{T} = \alpha^2 T/\kappa\rho \qquad (1)$$

There are various instruments to measure these parameters ($\alpha$, $\kappa$, and $\rho$) separately.[7–9] Measuring these parameters individually are time consuming. Also, repetitive heating and cooling may cause variation in the properties of thermoelectric materials.

There are few instruments available in the literature, that measure the thermoelectric parameters ($\alpha$, $\kappa$, and $\rho$), simultaneously above room temperature. However, for sample heating, some of them have used bulk heater[3,10,11] in which large area of the heater is exposed to the vacuum environment, which gives large radiation loss and high temperature difference between heating wire and sample hot side temperature[3,12]. Heat flux sensor is used to measure heat flow rate in thermal conductivity measurement[11]. These sensors are not suitable in high temperature region. *Muto et al.*[11] and *Amatya et al.*[3] have measured the relative Seebeck coefficient of the sample using low Seebeck coefficient reference materials (eg. Copper, Platinum, Niobium) as connecting wires.

*Kolb et al.*[10] used methodology for Seebeck coefficient measurement as described by *Boor et al.*[13], who assumed linear Seebeck coefficient of the thermocouple and its negative leg in the range of cold and hot side temperatures with condition that temperature difference should be low. Our group also used this methodology with the same approximation and fabricated Seebeck coefficient measurement setup.[14] Later, we removed this approximation and designed a very simple and lightweight sample holder.[15] For $\kappa$ measurement, *Zawilski et al.*[16] discussed parallel thermal conductance (PTC) technique to measure the rate of heat flow through the sample and performed measurement upto 300K. Further, *Dasgupta et al.*[17] used this technique and perform $\kappa$ measurement above room temperature. We simplified this technique and measured $\kappa$.[18] To measure $\alpha$, $\kappa$, and $\rho$, simultaneously, number of sensors and wires are required, which increases the complexity of the instrument. The number of signals required to measure at each datum, manual measurement of these signals required long lime where thermal stability is difficult to maintain.

In this work, we have fabricated a high temperature setup to measure *figure of merit* ($Z\bar{T}$). This setup facilitates the simultaneous measurement of Seebeck coefficient ($\alpha$), thermal conductivity ($\kappa$), and electrical resistivity ($\rho$) required to calculate $Z\bar{T}$. The setup design is simplified by using same thermocouples to perform temperature as well as voltage measurements. Limited components used in the sample holder further simplify the sample holder design and make it lightweight and user-friendly. Low heat loss is achieved by using a thin heater, which heats the sample cross section directly. Use of low thermal conductive gypsum insulator block and its optimized dimension further minimizes the heat loss. Use of commonly available materials for the fabrication of its components provides more accessibility for the users as any parts can be easily replaced in case any damage occurs. A dedicated program is built in the Python programming language to automate the whole measurement process. This setup is calibrated by using *p-type Bi*$_{0.36}$*Sb*$_{1.45}$*Te*$_3$ sample. The data collected are found to be in good agreement with the reported data.

## MEASUREMENT METHODOLOGY

The measurement of Seebeck coefficient is based on the differential method. We used the methodology suggested by *Boor et al.*[13], given below,

$$\alpha_{sample} = \frac{U_{neg}}{U_{pos} - U_{neg}} \alpha_{TC} + \alpha_{neg} \tag{2}$$

Where $U_{pos}$ and $U_{neg}$ are thermoelectric voltages measured using positive and negative legs of thermocouple wires, respectively. $\alpha_{TC}$ and $\alpha_{neg}$ are the Seebeck coefficients of the thermocouple and its negative leg, respectively. The main advantage of the above equation is that it does not require direct temperature measurement. The values of $\alpha_{TC}(T)$ and $\alpha_{neg}(T)$ are found out by using integral method discussed in the Ref. [15]. We used 36 SWG K-type PTFE shielded thermocouple. The temperature dependent values of $\alpha_{TC}(T)$ and $\alpha_{neg}(T)$ for K-type thermocouple are taken from Refs. [19] and [20], respectively.

Thermal conductivity ($\kappa$) is measured using one-dimensional Fourier's law of heat conduction, which can be written as

$$\kappa = \frac{\dot{Q}_s \Delta T}{Al} \tag{3}$$

Where $\dot{Q}_s$ and $\Delta T$ are the rate of heat flow through the sample and the temperature gradient across it. $l$ and A are sample thickness and its cross-sectional area, respectively. In this method, the accuracy of the $\kappa$ measurement mainly depends on the $\Delta T$ and $\dot{Q}_s$. The value of $\Delta T$ is obtained by subtracting cold side temperature ($T_c$) from hot side temperature ($T_h$).

The measurement of $\dot{Q}_s$ is difficult due to the undefined amount of heat losses by the means of conduction, convection, and radiation. In the measurement setup discussed in this work, the heat generated by the heater is mainly divided into two parts, one is through the sample and another is through the insulator block. The heat loss includes conductive heat flow through the insulator block, wires, and thermocouple connected at hot side. It also includes the radiation loss through the side walls of the sample and insulator block.

To measure heat loss, two cycles of the measurements are required. In the first cycle, measurement is performed by running the instrument without the sample. This heat loss ($\dot{Q}_h$) value includes the conduction and radiation loss through the insulator block, radiation loss through hot side copper block. This heat loss also includes the heat flow through the wires and thermocouple at hot side due to the cold finger effect.[21] This heat loss data along with hot side temperature relative to the vacuum chamber temperature ($T_{hr}$) have been recorded to find out the heat loss value during κ measurement using the PTC technique discussed in the Ref. [18].

The heat loss value measured in the first measurement cycle does not include the radiation loss through side walls of the sample ($\dot{Q}_{ls}$). To find this loss, we perform another heat loss measurement. In this measurement cycle, the sample is placed over the hot side copper block without fixing cold side copper block over it. Power supplied to the heater with $T_{hr}$ value are recorded. $\dot{Q}_{ls}$ can be obtained by subtracting the power supplied to the heater in the first cycle from the second cycle. Here, we assumed that at the same $T_{hr}$ value, the heat flow through the insulator block, hot side wires and thermocouple are same. This obtained heat loss through sample with $T_{hr}$ data is fitted with a polynomial function ($\dot{Q}_{ls}(T)$).

The temperature profile of the sample during heat loss measurement and thermal conductivity measurement are shown in the Fig. 1(a), and 1(b), respectively. During heat loss measurement with the sample, the temperature throughout the sample is assumed to be same, while during thermal conductivity measurement a temperature gradient will be generated across the thickness of the sample. To find the rate of heat loss through the sample surface ($\dot{Q}_{ls}$) during thermal conductivity (κ) measurement, we use an integration method given below,

$$\dot{Q}_{ls}(T_{cr}, T_{hr}) = \frac{1}{\Delta T} \int_{T_{cr}}^{T_{hr}} \dot{Q}_{ls}(T) dT \qquad (4)$$

Where $T_{cr}$ is cold side temperature relative to the vacuum chamber temperature. This gives the effective heat loss through the sample surface.

Now, the net heat loss during measurement can be found out by using given equation,

$$\dot{Q}_l(T_r) = \dot{Q}_{li}(T_r) + \dot{Q}_{ls}(T_{hr}, T_{cr}) \qquad (5)$$

During κ measurement, the net heat flow through the sample is obtained by subtracting $\dot{Q}_l$ from the power supplied to the heater at that particular $T_{hr}$ value.

Electrical resistivity (ρ) is measured using four-wire lead arrangement.[22] Nickel wires are used to feed current (I) to the sample. For this, a nickel wire is fixed in the one hole of each copper block. Negative legs of both thermocouples are used to measure voltage ($V_1$) across the sample. This measured voltage also includes thermoelectric voltage ($V_{TE}$) across the sample.[23] The finite temperature gradient across the sample may create large $V_{TE}$, which leads to large uncertainty in the electrical resistivity measurement. To nullify this thermoelectric voltage, we reverse the direction of the current (-I) and measure voltage ($V_2$) again. Mathematically $V_1$ and $V_2$ can be written as,

$$V_1 = V_{IR} + V_{TE} \tag{6}$$

$$V_2 = -V_{IR} + V_{TE} \tag{7}$$

Where $V_{IR}$ is the resistive voltage due to the flow of current I through the sample. By using above two equations, $V_{IR}$ can be written as,

$$V_{IR} = \frac{V_1 - V_2}{2} \tag{8}$$

Due to very high electrical conductivity of copper block, as compared to that of the conventional thermoelectric samples, the current fed for electrical resistivity measurement spread over the full base face of the sample.[10] Use of high electrical conductive GaSn liquid metal at the interface surfaces minimizes the electrical resistance between sample and copper blocks largely, so it can be neglected.[3]

The electrical resistivity (ρ) of a sample with uniform cross section can be obtained by the equation given below[10],

$$\rho = \frac{AV_{IR}}{Il} \tag{9}$$

Due to Peltier effect, the flow of current through the sample can create an additional temperature difference across the sample. Error due to the Peltier effect still exists even after averaging of measured voltages by reversing the current. Time taken for Peltier heat to diffuse from one contact to distance $x$ is proportional to $x^2/D$, where D is thermal diffusivity of thermoelectric material.[24] Typically it lies between 1.1s-1.35s for common thermoelectric materials.[25] To nullify this effect current is reversed in less than 0.2 s so that error due the Peltier effect is very small and can be neglected.[26]

After the measuring of $\alpha$, $\kappa$, and $\rho$, figure of merit ($Z\bar{T}$) value is calculated at each datum using the equation 1 discussed in the introduction.

## MEASUREMENT SETUP

The actual photograph of the setup used for $Z\bar{T}$ measurement is shown in Fig. 2. It's different components are represented by numbers. High thermal and electrical conductive copper blocks, *1*, are used to sandwich the sample. The GaSn liquid metal is used at the interface surfaces between copper blocks and sample. High boiling temperature (~1800 K) of GaSn makes it suitable for high temperature application. It has high thermal and electrical conductivity as compared to the conventional thermoelectric materials. Due to this, its thin layer at the interface surfaces adds very small thermal and electrical resistances in the measurement, which can be ignored. Each copper block has two holes at the sidewalls, which are used to fix thermocouple, *2*, and conducting wire, *3*. This conducting wire is used to feed current, which is required for the electrical resistivity measurement. 36 SWG K-type PTFE shielded thermocouples are used for temperature and voltage measurement. 12 Ω thin heater, *4*, is used to heat the sample. It is made by winding 40 SWG kanthal wire over a thin mica sheet. Both ends of kanthal wire are welded with copper wires. To measure the power delivered to the heater accurately, we implemented 4-wire technique at the electrical connector inside the chamber. This thin heater is placed over an insulator block, *5*, made of low thermal conductive gypsum board to minimize the heat loss.

Again, heat loss is minimized by optimizing insulator cross-section and its thickness. This insulator block is supported by a rectangular brass bar, *6*, by using high temperature cement. Thin mica sheet, *7*, is also used above the cold side copper block to insulate sample electrically. Good surface contact between sample and copper block is ensured by screwing threaded SS rod, *8*. The round tip of this SS rod ensures self-aligning of the cold side copper block over the sample surface and maintains almost constant pressure throughout the cross section of the sample. Another rectangular brass bar, *9*, is used to hold this SS rod. Both rectangular brass bars have been fixed over two separate SS rods, *10*, supported by the SS flange. PT-100 RTD is connected to the electrical connector to monitor the vacuum chamber temperature. The temperature of the measurement area during measurement is fixed to 300K. SS flange with sample holder assembly are put over the vacuum chamber, which is connected to the vacuum pump. Rotary vacuum pump is used to create a vacuum inside the chamber upto a level of ∼ $8 \times 10^{-3}$ mbar.

Digital multimeter (DMM) with multi-channel scanner card is used to measure various signals. Dual channel sourcemeter (SMU) is used to supply power to the heater and feed current to the sample for electrical resistivity measurement. DMM and SMU are interfaced with the computer using GPIB interface. The whole measurement process has been automated by using Python. This program uses pyVisa, Numpy, and Scipy library functions.[27]

**RESULTS AND DISCUSSIONS**

Low thermal conductive gypsum insulator block minimizes the conductive heat loss through the insulator block. The heat flow through the wires and thermocouple attached to the heater and hot side copper block also cause heat loss due to the cold finger effect.[21]

The heat loss data measured with and without $Bi_{0.36}Sb_{1.45}Te_3$ sample are shown in the Fig. 3. Very small surface area (< 9$mm^2$) of $Bi_{0.36}Sb_{1.45}Te_3$ sample causes a very low radiation loss compared to the heat loss through the insulator block at that $T_{hr}$ value. This creates a very small gap between both heat loss data, which are difficult to obtain experimentally. It shows that the proposed methodology to measure radiation loss through the sidewalls of the sample will be useful in case of the sample with larger sidewalls area. Currently, Heusler alloys are the potential thermoelectric candidates and having wide thermal conductivity range (upto 16 W/m-K at room temperature).[28–30] The characterization of these samples over this setup requires larger thickness to get reasonable temperature gradient across it. This large thickness results in the large sidewalls area, where the considerable radiation loss happens.

To verify the instrument, $Z\bar{T}$ measurement has been performed over p-type bead of $Bi_{0.36}Sb_{1.45}Te_3$ sample from mean temperature ($\bar{T}$) of 305K to 508K. This sample has been extracted from the commercially available thermoelectric module (TEC1–12706). To obtain the composition of the sample, energy-dispersive X-ray spectroscopy is performed. The cross-sectional area and thickness of the sample are 1.96 $mm^2$ and 1.6 mm, respectively.

The temperature gradient generated across this sample with mean temperature are shown in the Fig. 4. At the start of the measurement, the value of $\Delta T$ is 2 K. It increases almost linearly till ~400K. At this temperature, the slope of the curve changes and again $\Delta T$ increases almost linearly and reaches to 95K at the end of the measurement.

The measured data with error bar and reported data by *Ma et al.*[31] and *Poudel et al.*[32] of $\alpha$, $\kappa$, $\rho$, and $Z\bar{T}$ have been shown in the Fig 5, 6, 7, and 8, respectively. At $\bar{T}$ = 305K, the Seebeck coefficient of the sample is 200 $\mu$V/K, it increases with increase in the temperature till 350K and reaches to 231 $\mu$V/K. After this temperature, it starts decreasing with increase in the temperature till the end of the measurement and reaches to 125 $\mu$V/K. At the start of the measurement, the value of thermal conductivity is ~1.4 W/m-K. Change in its value till 370K is very small. After this temperature, the thermal conductivity increases almost linearly till the end of the measurement and reaches to 2.64W/m-K.

Initially, the electrical resistivity increases almost linearly till 410K and changes from 1.28 to 2.18. After this temperature, it increases slowly and reaches to 2.5 at 510K. The figure of merit ($Z\bar{T}$) value at 305K is ~0.85. Till 340K its value is almost constant and then decreases with increase in the temperature and reaches to 0.12 at 510K.

Measured data have been compared with the data reported by *Ma et al.*[31] and *Poudel et al.*[32]. Data reported in the both literature are on $Bi_{0.5}Sb_{1.5}Te_3$ commercial ingot in which sample composition is verified by comparing the XRDs data of their samples with the standard XRD data of $Bi_{0.5}Sb_{1.5}Te$3. From 305K to 345K, Seebeck coefficient data match closely with the data reported by *Ma et al.*[31] and show a small deviation with the data reported by *Poudel et al.*[32]. With the increase in temperature, the deviation in the measured data increases, whereas deviation between both reported data also increases. At 510K, measured data show a deviation of 20 $\mu$V/K with the data reported by *Ma et al.*[31] while a deviation of 22 $\mu$V/K was observed between both reported data.

From 305K to 370K, measured thermal conductivity data match closely with the both reported data. The deviation in measured data starts increasing after this temperature. The maximum deviation in measured data with the data reported by *Ma et al.*[31] is 0.4W/m-K at the end of the measurement, while a deviation of 0.2W/m-k has been observed among the both reported data.

At 305K, the measured electrical resistivity value shows a deviation of 0.2 mΩ-cm with the data reported by *Ma et al.*[31] While a deviation of 0.05 mΩ-cm has been observed between both the reported data. The deviation increases gradually and reaches to 0.3 mΩ-cm at the end of the measurement. The deviation of 0.2 mΩ-cm has been observed between both reported data at 525K.

At 305K, measured $Z\bar{T}$ value shows a deviation of 0.15 from the reported data by *Ma et al.*[31] At this temperature, a deviation of 0.05 is observed between both the reported data. The deviation in measured data increases gradually and reaches to 0.24 at 450K and then starts decreasing. A deviation of 0.1 is observed at 508K. At this temperature, a deviation of 0.05 is between both the reported data.

The composition of measured sample is somewhat different from the samples used in the reported data. This may be the main reason for these observed deviations. In the both reported data, the temperature gradient generated across the sample during measurement and the measurement methodology had not discussed. We performed measurement in which temperature gradient increases almost linearly, resulting in large $\Delta T$ at high temperature. This may also be the reason for deviation at higher temperatures. The various available literature shows that the sample preparation technique is very important and yields different properties even with the same sample composition.

The successive measurements have been performed to verify reproducibility and repeatability of the instrument on the same sample by remounting the sample for each measurement. The maximum error in the Seebeck coefficient has been observed below 1%. In thermal conductivity measurement, a maximum error of 0.2 W/m-K has been observed at the start of the measurement. This is mainly due to the very small heat flow through the sample. This error decreases with increase in the temperature due to the increase in the heat flow. The maximum error of ~5% has been observed in the $\rho$ and $Z\bar{T}$ measurement.

## CONCLUSION

Fabrication of high temperature Figure of merit ($Z\bar{T}$) measurement setup has been reported. The designed setup facilitates the simultaneous measurement of Seebeck coefficient ($\alpha$), thermal conductivity ($\kappa$), and electrical resistivity ($\rho$) required to find $Z\bar{T}$. The most simplified version of the setup is achieved due to the measurement of temperature as well voltage by using same thermocouples. Setup design has been further simplified by using limited components and thin heater. Use of thin heater minimizes the heat loss, which is very important for $\kappa$ measurement. Further, the heat loss is minimized by using low thermal conductive gypsum insulating block and its optimized dimension. The power delivered to the heater is measured accurately by using the 4-wire technique. Low cost and commonly available materials used in the fabrication of various components make it more accessible to the user as any part can be easily replaced in case any damage occurs. A dedicated program based on the Python is built to automate the whole measurement process. $Bi_{0.36}Sb_{1.45}Te_3$ based *p-type* sample is used to calibrate this setup. The data collected are found to be in good agreement with the reported data.

## ACKNOWLEDGEMENTS

The authors acknowledge workshop staffs for their support in the fabrication of various components of vacuum chamber and sample holder.

FIG. 1. Temperature profile of sample during (a) heat loss measurement, and (b) thermal conductivity measurement where, $T_c$ and $T_h$ are cold side and hot side temperature, respectively.

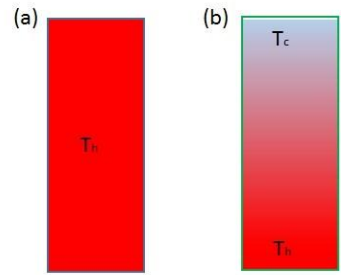

FIG. 2. The actual photograph of sample holder: (1) copper blocks, (2) thermocouple, (3) conducting wire, (4) thin heater, (5) insulator block, (6) rectangular brass bar, (7) thin mica sheet, (8) threaded SS rod, (9) rectangular brass bar, and (10) SS rods.

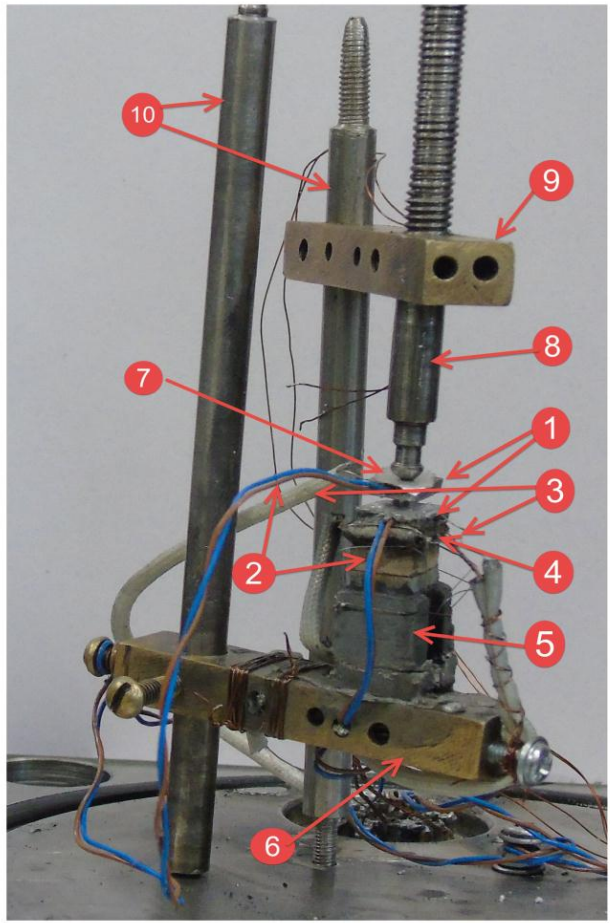

FIG. 3. The heat loss without and with sample at different $T_{hr}$ (hot side temperature with respect to reference temperature)

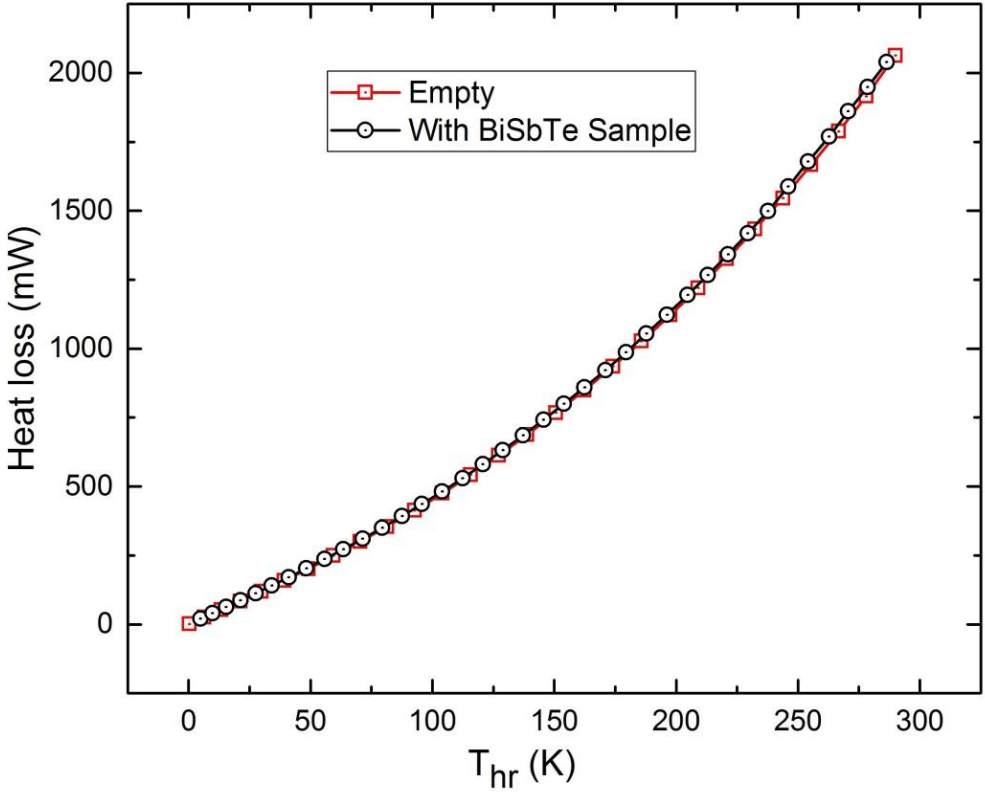

FIG. 4. Variation in the temperature gradient (Δ$T$) with the mean temperature ($\bar{T}$)

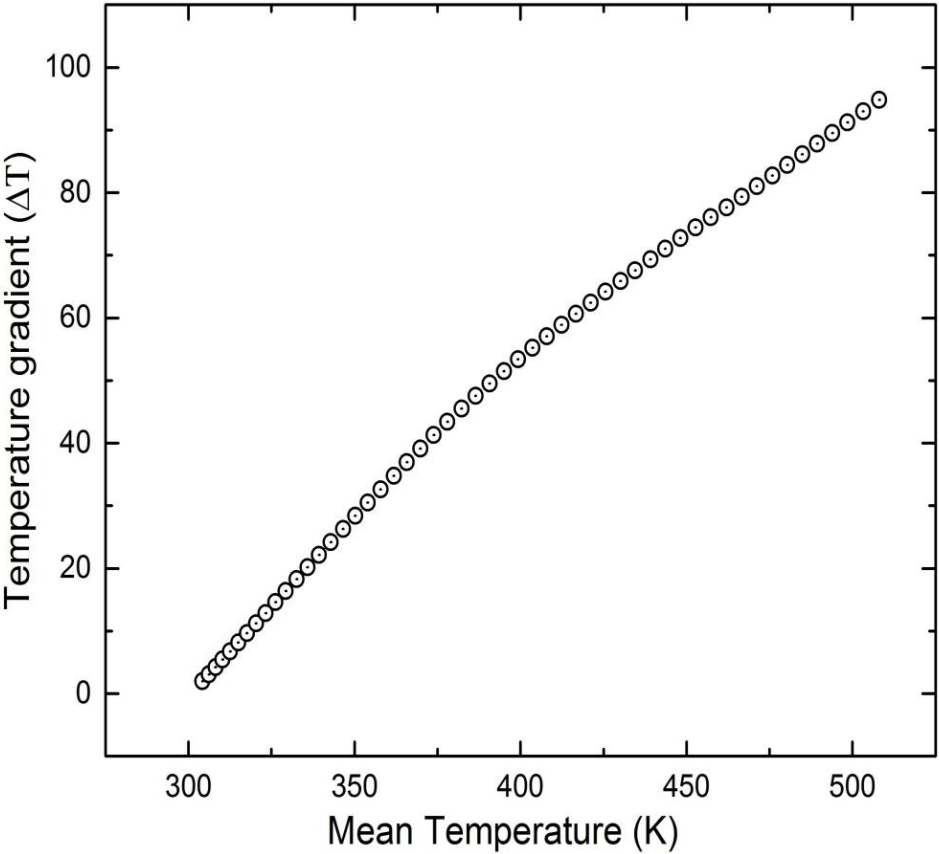

FIG. 5. Seebeck coefficient ($\alpha$) with mean temperature ($\bar{T}$) of *BiSbTe* sample measured using this setup (open circle), data reported by *Ma et al.*[31] (open triangle), and *Poudel et al.*[32] (open box).

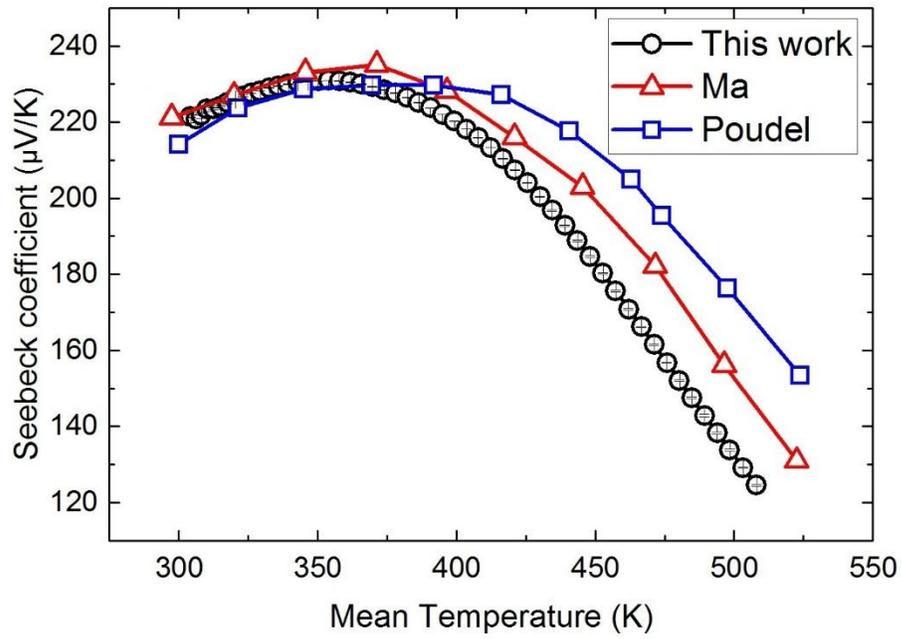

FIG. 6. thermal conductivity ($\kappa$) with mean temperature ($\bar{T}$) of *BiSbTe* sample measured using this setup (open circle), data reported by *Ma et al.*[31] (open triangle), and *Poudel et al.*[32] (open box).

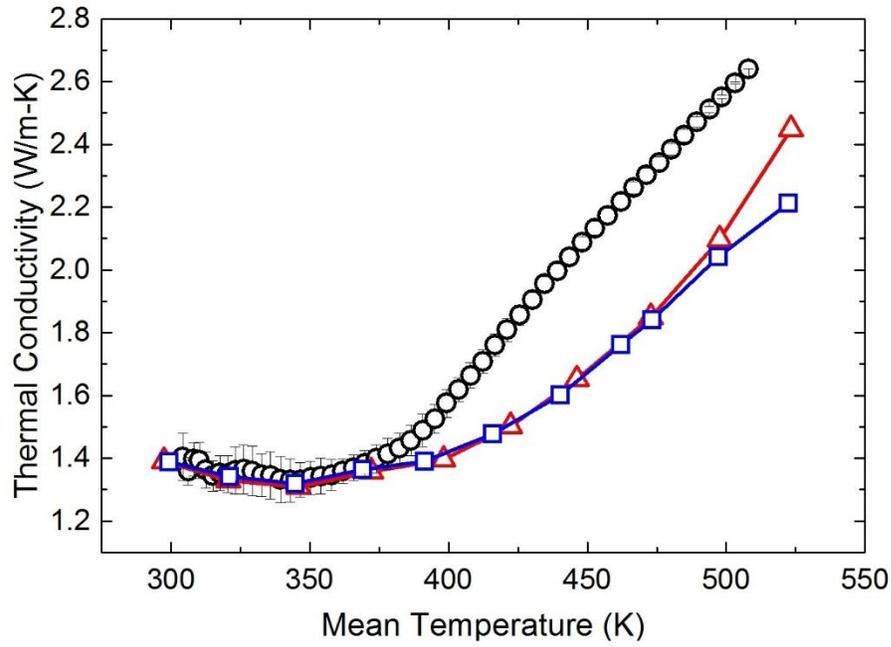

FIG. 7. (c) electrical resistivity ($\rho$) with mean temperature ($\bar{T}$) of *BiSbTe* sample measured using this setup (open circle), data reported by *Ma et al.*[31] (open triangle), and *Poudel et al.*[32] (open box).

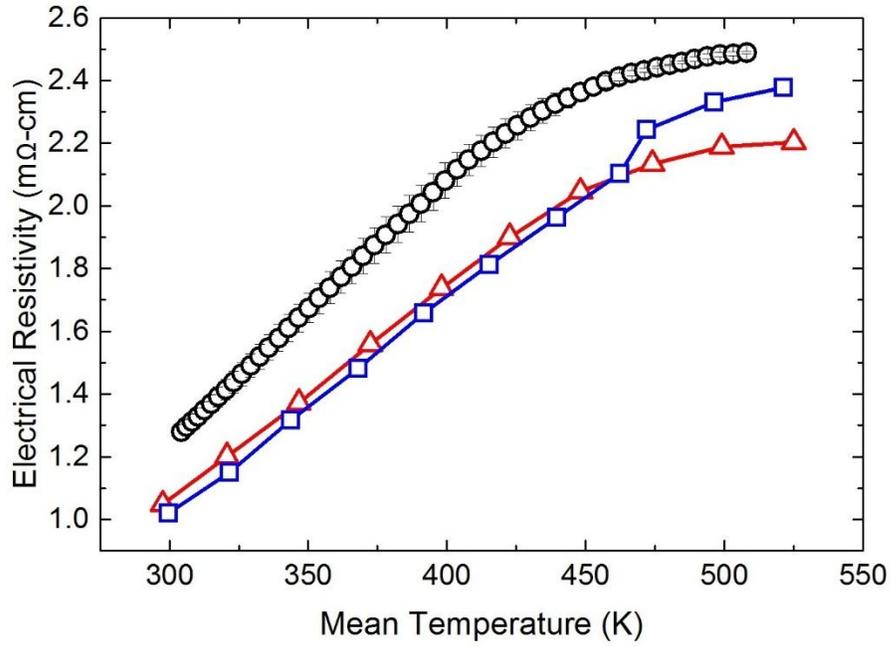

FIG. 8. figure of merit ($Z\bar{T}$) with mean temperature ($\bar{T}$) of *BiSbTe* sample measured using this setup (open circle), data reported by *Ma et al.*[31] (open triangle), and *Poudel et al.*[32] (open box).

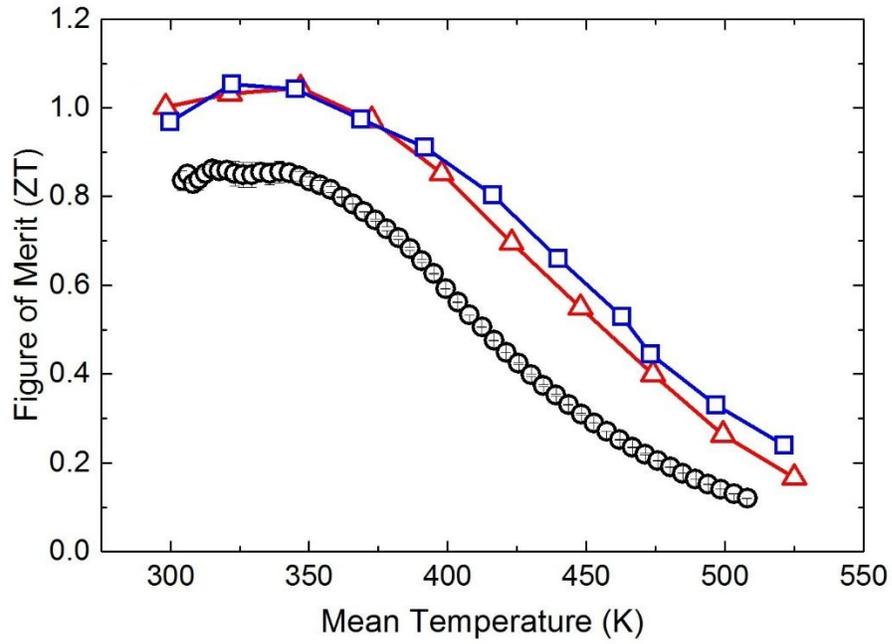